\documentclass[aps,pra,twocolumn,showpacs]{revtex4}
\usepackage{graphicx,graphics,psfrag,amsmath,calc,amssymb}

\begin{document}

\title{BCS-BEC Crossover of a Quasi-two-dimensional Fermi Gas: the
Significance of Dressed Molecules}
\author{Wei Zhang, G.-D. Lin, and L.-M. Duan}
\affiliation{FOCUS center and MCTP, Department of Physics,
University of Michigan, Ann Arbor, MI 48109}
\date{\today}

\begin{abstract}
We study the crossover of a quasi-two-dimensional Fermi gas trapped
in the radial plane from the Bardeen-Cooper-Schrieffer (BCS) regime
to the Bose-Einstein condensation (BEC) regime by crossing a wide
Feshbach resonance. We consider two effective two-dimensional
Hamiltonians within the mean-field level, and calculate the zero
temperature cloud size and number density distribution. For model 1
Hamiltonian with renormalized atom-atom interaction, we observe a
constant cloud size for arbitrary detunings. For model 2 Hamiltonian
with renormalized interactions between atoms and dressed molecules,
the cloud size deceases from BCS to BEC side, which is consistent
with the picture of BCS-BEC crossover. This qualitative discrepancy
between the two models indicates that the inclusion of dressed
molecules is essential for a mean-field description of
quasi-two-dimensional Fermi systems, especially on the BEC side of
the Feshbach resonance.
\end{abstract}

\maketitle



\section{introduction}
\label{sec:introduction}

The interest on low-dimensional Fermi systems has been recently
reinvoked by the experimental developments of cooling and trapping
atoms in optical lattices~\cite{1,2,3} and on atom chips~\cite{4}.
With the aid of tuning an external magnetic field through a Feshbach
resonance, these techniques provide a fascinating possibility of
creating quasi-low-dimensional Fermi systems with a controllable
fermion-fermion interaction. In particular, the interaction between
fermions can be tuned from a Bardeen-Cooper-Schrieffer (BCS) limit
to a Bose-Einstein condensation (BEC) limit, such that the BCS-BEC
crossover can be studied in quasi low dimensions. The BCS-BEC
crossover has been extensively studied in three-dimensional (3D)
Fermi systems, where a single-channel model~\cite{5} and a
two-channel model~\cite{6} are both applied to give a consistent
description around a wide Feshbach resonance. This agreement between
single- and two-channel models is rooted from the fact that the
closed-channel (Feshbach molecule) population is negligible near a
wide resonance, so it will not cause any significant difference by
taking the molecules into account (as in the two-channel model) or
completely neglecting them (as in the single-channel model). The
BCS-BEC crossover of a uniform two-dimensional (2D) Fermi system has
also been considered in connection with high-$T_{c}$
superconductors~\cite{7}, where an effective 2D Hamiltonian with
renormalized fermion-fermion interaction is employed.

In this paper, we study the BCS-BEC crossover in a quasi-2D Fermi
gas, first using an effective 2D Hamiltonian with renormalized
atom-atom interaction (model 1) \cite{petrov-00, model-1}, and then a more
general model with renormalized interaction between atoms and
dressed molecules (model 2)~\cite{kestner-07}. The dressed
molecules mainly come from population of atoms in the excited
levels along the strongly confined axial direction near a Feshbach
resonance~\cite{kestner-07,duan-05}. When considering the effect
of a weak harmonic trap in the two loosely confined dimensions
under the local density approximation (LDA), we adapt the
mean-field (MF) treatment to calculate the zero temperature cloud
size and number density distribution in the radial plane. We find
a significant difference between the two models. By using model 2,
we show that the cloud size decreases from the limiting value of a
weakly interacting Fermi gas as one moves from the BCS to the BEC
side of the Feshbach resonance, and approaches to the limiting
value of a weakly interacting Bose gas in the BEC limit. This
behavior is a signature of the BCS-BEC crossover in quasi two
dimensions. On the contrary, model 1 fails to describe this
crossover behavior, but predicts a constant cloud size and
identical density profile for all magnetic field detunings. This
discrepancy implies that the MF results given by model 1 is
unreliable, even at a qualitative level. Given this qualitative
discrepancy and the problem associated with model 1 for
description of the two-body ground state of the system
\cite{kestner-06}, it is likely that the oversimplification is
rooted in the model itself instead of the mean-field
approximation.

The quasi-2D geometry can be realized by arranging a one-dimensional
(1D) optical lattice along the axial ($z$) direction and a weak
harmonic trapping potential in the radial ($x$-$y$) plane, such that
fermions are strongly confined along the $z$ direction and form a
series of quasi-2D pancake-shaped clouds~\cite{3}. Each such
pancake-shaped cloud can be considered as a quasi-2D Fermi gas when
the axial confinement is strong enough to turn off inter-cloud
tunneling. The strong anisotropy of trapping potentials introduces
two different orders of energy scales, with one characterized by
$\hbar \omega_{z}$ and the other by $\hbar \omega_{\perp}$, where
$\omega_{z}$ ($\omega_{\perp }$) are the trapping frequencies in the
axial (radial) directions. The separation of these two energy scales
($\omega_{z}\gg \omega_{\perp }$) allows us to first deal with the
axial degrees of freedom and derive an effective 2D Hamiltonian, and
leave the radial degrees of freedom for later treatment.


\section{Model 1 with renormalized atom-atom interaction}
\label{sec:model1}

The effective 2D Hamiltonian for model 1 is obtained by assuming
that the renormailied atom-atom interaction can be characterized
with an effective 2D scattering length, with the latter derived
from the exact two-body scattering physics \cite{petrov-00, model-1}. Thus,
for a wide Feshbach resonance where the Feshbach-molecule
population is negligible, we can write down an
effective Hamiltonian only in terms of 2D fermionic operators $a_{\mathbf{k}%
,\sigma }$ and $a_{\mathbf{k},\sigma }^{\dagger }$, with (pseudo) spin $%
\sigma $ and transverse momentum $\mathbf{k}=\left(
k_{x},k_{y}\right) $. The model 1 Hamiltonian thus takes the
form~\cite{petrov-00, model-1,7}
\begin{eqnarray}
H_{1} &=& \sum_{\mathbf{k},\sigma }\left( \epsilon _{\mathbf{k}}-\mu \right)
a_{\mathbf{k},\sigma }^{\dagger }a_{\mathbf{k},\sigma }  \notag
\label{eq:sc-Hamiltonian} \\
&+&
\frac{V_{1}^{\text{eff}}}{L^{2}}\sum_{\mathbf{k},\mathbf{k^{\prime}},\mathbf{q}}
a_{\mathbf{k},\uparrow}^{\dagger} a_{-\mathbf{k}+\mathbf{q},\downarrow}^{\dagger}
a_{\mathbf{k^{\prime}},\downarrow}a_{-\mathbf{k^{\prime}}+\mathbf{q},\uparrow},
\end{eqnarray}
where $\epsilon_{\mathbf{k}}=\hbar^{2} k^{2}/(2m)$ is the 2D
dispersion relation of fermions with mass $m$, $\mu$ is the chemical
potential, and $L^{2}$ is the quantization area. The bare parameter
$V_{1}^{\text{eff}}$ is connected with the physical one $V_{1p}^{\text{eff}}$
through the 2D renormalization relation
$\left[ V_{1}^{\text{eff}}\right]^{-1}=\left[ V_{1p}^{\text{eff}}\right]^{-1}
-L^{-2}\sum_{\mathbf{k}}\left( 2 \epsilon_{\mathbf{k}}+\hbar \omega_{z}\right)^{-1}$
($\hbar \omega _{z}$ is from the zero-point energy), and
$V_{1p}^{\text{eff}}=V_{1p}^{\text{eff}}(a_{s},a_{z})$ depends on the
3D scattering length $a_{s}$ and the characteristic length scale for axial motion
$a_{z}\equiv \sqrt{\hbar /(m\omega_{z})}$ with the expression given in
Ref.~\cite{petrov-00, model-1, kestner-07}. Notice that the chemical potential $\mu$
can be a function of the radial coordinate $\mathbf{r}=(x,y)$ under LDA.
In the following discussion, we choose $\hbar \omega_{z}$ as the energy unit
so that $\mu $, $V_{1}^{\text{eff}}$, and $\epsilon_{\mathbf{k}}=a_{z}^{2} k^2/2$
become dimensionless.

By introducing a BCS order parameter (also dimensionless in unit of
$\hbar\omega_{z}$) $\Delta \equiv (V_{1}^{\text{eff}}/L^{2})\sum_{\mathbf{k}}
\left\langle a_{\mathbf{k},\downarrow }a_{-\mathbf{k},\uparrow }\right\rangle$,
we get the zero temperature thermodynamic potential density
\begin{equation}
\Omega =-\frac{\Delta ^{2}}{V_{1}^{\text{eff}}}+\frac{1}{L^{2}}
\sum_{\mathbf{k}}\left( \epsilon_{\mathbf{k}}-\mu -E_{\mathbf{k}}\right),
\end{equation}
where $E_{\mathbf{k}}=\sqrt{(\epsilon_{\mathbf{k}}-\mu )^{2}+\Delta^{2}}$
is the quasi-particle excitation spectrum. The ultraviolet divergence of the
summation over $\mathbf{k}$ cancels with the renormalization term in
$\left[V_{1}^{\text{eff}}\right]^{-1}$. The gap and number equations can be
obtained respectively from $\partial \Omega /\partial \Delta ^{2}=0$ and
$n=-\partial \Omega /\partial \mu $ ($n=N/L^{2}$ is the density of particles),
leading to
\begin{eqnarray}
\frac{1}{V_{1p}^{\text{eff}}(a_{s},a_{z})} &=&\frac{\ln
\left( -\mu +\sqrt{\mu ^{2}+\Delta ^{2}}\right) }{4\pi a_{z}^{2}},
\label{eq:gap1} \\
n &=&\frac{\mu +\sqrt{\mu ^{2}+\Delta ^{2}}}{2\pi a_{z}^{2}}.
\label{eq:number1}
\end{eqnarray}
Notice that Eq. (\ref{eq:gap1}) can be rewritten as
$F(a_{s},a_{z})=-\mu +\sqrt{\mu^{2}+\Delta ^{2}}$, where the function
$F$ absorbs all the dependence on $a_{s}$ and $a_{z}$.
Thus, by substituting this expression into Eq. (\ref{eq:number1}),
we get a closed form for the number equation,
\begin{equation}
n=\frac{1}{\pi a_{z}^{2}}\left[ \frac{F(a_{s},a_{z})}{2}+\mu \right] .
\label{eq:number3}
\end{equation}

Now we take into account the harmonic trapping potential
$U(\mathbf{r})=(\omega_\perp/\omega_z)^2 r^2 /(2 a_z^2)$ in the radial plane
by writing down the position dependent chemical potential
$\mu(\mathbf{r})=\mu_0-U(\mathbf{r})$, where $\mu_0$ is the chemical potential
at the trap center. It can be easily shown that the spacial density profile
is now a parabola, $n(\mathbf{r})=(\omega_\perp/\omega_z)^2
(R_{\mathrm{TF}}^2-r^2)/(2\pi a_z^4)$, with the Thomas-Fermi cloud size
$R_{\mathrm{TF}}=\sqrt{2\mu_0} a_z (\omega_z/\omega_\perp)$. By assigning the
condition that the total number of particles in the trap is fixed by
$N=\int n(\mathbf{r}) d^2 \mathbf{r}$, the cloud size takes the constant value
$R_{\mathrm{TF}}=R_{\mathrm{BCS}} \equiv \sqrt{2 \omega_z/\omega_\perp} (N)^{1/4} a_z$,
which is independent on the 3D scattering length $a_s$. In fact, as one varies the
scattering length $a_s$, the chemical potential at the trap center $\mu_0$
is adjusted accordingly such that the identical density profile is maintained.

This result of a constant cloud size is obviously inconsistent with
the picture of a BCS-BEC crossover in quasi two dimensions. In fact,
in a typical experiment with $a_z$ ($\sim \mu$m) much greater than
the interatomic interaction potential $R_e$ ($\sim $nm), the
scattering of atoms in this quasi-2D geometry is still 3D in nature.
In particular, fermions will form tightly bound pairs on the BEC
side of the Feshbach resonance as they do in 3D. Thus, in the BEC
limit when fermion pair size $a_{\rm pair} \ll a_z$ and binding
energy $\vert E_b \vert \gg \hbar \omega_z$, the system essentially
behaves like a weakly interacting gas of point-like bosons, for
which one would expect a vanishing small cloud size in the loosely
confined radial plane~\cite{petrov-00, posazhennikova-06}.

The MF result of a finite cloud size in the BEC limit from model 1
indicates a finite interaction strength between paired fermions, no
matter how small they are in size. This statement can be extracted
directly from the number equation (\ref{eq:number3}), which can be
written in the form $\mu=n\pi a_{z}^{2}-F(a_{s},a_{z})/2$. In the
BEC limit, the second term on the right-hand side denotes one half
of the binding energy, while the first term indicates a finite
interaction energy per fermion pair since it is proportional to the
number density. As a comparison, the actual equation of state for
fermion pairs one should expect must take the form as for a quasi-2D
Bose gas in the weakly interacting limit~\cite{petrov-00}
\begin{equation}
\label{eq:eq-of-state}
\mu_B \approx 3 n a_z a_s,
\end{equation}
in which case the quasi-2D gas is treated as a 3D condensate with
the ground state harmonic oscillator wave function in the
$z$-direction.

The interaction strength between paired fermions can also be derived
by writing down a Bose representation for this system, where the
fermionic degrees of freedom are integrated out in the BEC
limit~\cite{tokatly-04}. This Bose representation leads to a
two-dimensional effective Hamiltonian for bosonic field $\phi({\bf
r})$,
\begin{equation}
\label{eq:eff-H}
H_{\rm eff} = \int d{\bf r}
\left[
\phi^\dagger({\bf r}) \left( - \frac{\hbar^2 \nabla^2 }{4m}
+ 2 U ({\bf r}) \right) \phi({\bf r})
- \frac{g_2}{2} \vert\phi({\bf r})\vert^4
\right],
\end{equation}
where the quartic term characterizes the bosonic interaction. Within
the stationary phase approximation, the interaction strength $g_2$
is calculated by the leading diagram of a four-fermion process with
four external boson lines and four internal fermion propagators,
leading to~\cite{tokatly-04}
\begin{equation}
\label{eq:four-body}
g_2=2 \sum_{{\bf p},\omega} \Lambda_0^4({\bf p})
G_0^2({\bf p}, \omega) G_0^2(-{\bf p}, -\omega).
\end{equation}
Here, $\Lambda_0 = (-p^2/m + \vert E_{b} \vert) \chi_0({\bf p})$ is
the boson-fermion vertex, $\chi_0({\bf p})$ is the Fourier transform
of the relative wave function $\chi_0({\bf r})$ of two colliding
fermions in the $s$-wave channel, and $G_0({\bf p}, \omega) =
(i\omega - p^2/2m - \vert E_b \vert/2)^{-1}$ is the free propagator
for fermions. After summing over momentum ${\bf p}$ and Matsubara
frequency $\omega$, we can directly show that $g_2$ indeed takes a
constant value, being independent of the binding energy $\vert E_b
\vert$ of paired fermions and hence the 3D scattering parameter
$a_s$. Thus, we conclude that the MF theory based on model 1 fails
to recover the picture of a weakly interacting Bose gas of paired
fermions in the BEC limit, and can not be directly applied to
describe the BCS-BEC crossover in quasi two dimensions.


\section{Model 2 with inclusion of dressed molecules}
\label{sec:model2}

Having shown the problem associated with model 1, next we consider
model 2 by taking into account the axially excited states via
inclusion of dressed molecules. As derived in
Ref.~\cite{kestner-07}, the effective 2D Hamiltonian takes the form
(also in unit of $\hbar \omega_{z}$),
\begin{eqnarray}
H_{2} &=& \sum_{\mathbf{k},\sigma }\left( \epsilon _{\mathbf{k}}-\mu \right)
a_{\mathbf{k},\sigma }^{\dagger }a_{\mathbf{k},\sigma }
+\sum_{\mathbf{q}}\left( \frac{\epsilon _{\mathbf{q}}}{2}+\lambda_{b}-2\mu \right)
d_{\mathbf{q}}^{\dagger }d_{\mathbf{q}}
\notag  \label{eq:tc-Hamiltonian} \\
&+& \frac{\alpha_{b}}{L}\sum_{\mathbf{k},\mathbf{q}}\left( a_{\mathbf{k},\uparrow }^{\dagger }
a_{-\mathbf{k}+\mathbf{q},\downarrow }^{\dagger }d_{\mathbf{q}}+h.c.\right)
\notag \\
&+&\frac{V_{b}}{L^{2}}\sum_{\mathbf{k},\mathbf{k^{\prime }},\mathbf{q}}
a_{\mathbf{k},\uparrow }^{\dagger }a_{-\mathbf{k}+\mathbf{q},\downarrow}^{\dagger}
a_{-\mathbf{k^{\prime }},\downarrow }a_{\mathbf{k^{\prime }}+\mathbf{q},\uparrow },
\end{eqnarray}
where $d_{\mathbf{q}}^{\dagger }$ ($d_{\mathbf{q}}$) denotes the creation
(annihilation) operator for dressed molecules with radial momentum $\mathbf{q}$,
and $\lambda _{b}$, $\alpha _{b}$, and $V_{b}$ are the 2D effective bare detuning,
atom-molecule coupling rate, and background interaction, respectively.
These parameters can be related to the corresponding 3D parameters by matching
the two-body physics~\cite{kestner-07}. By introducing the order parameter
$\Delta \equiv \alpha_{b}\left\langle d_{0}\right\rangle /L+(V_{b}/L^{2})
\sum_{\mathbf{k}}\left\langle a_{\mathbf{k},\downarrow }a_{-\mathbf{k},\uparrow}\right\rangle $,
we obtain the mean-field gap and number equations,
\begin{eqnarray}
&&
\frac{1}{V_{2p}^{\text{eff}}(2\mu)}=
\frac{\ln \left(-\mu+\sqrt{\mu^{2}+\Delta^{2}}\right)}{4\pi a_{z}^{2}},
\label{eq:gap2} \\
&&
n=\frac{\mu +\sqrt{\mu^{2}+\Delta^{2}}}{2\pi a_{z}^{2}}+2\Delta^{2}
\frac{\partial \lbrack V_{2,\mathrm{p}}^{\text{eff}}(x)]^{-1}}{\partial x}\Bigg\vert_{x=2\mu},
\label{eq:number2}
\end{eqnarray}
where the inverse of effective interaction is connected with the 3D physical
parameters through \cite{kestner-07}
\begin{eqnarray}
\left[ V_{2p}^{\text{eff}}(x)\right] ^{-1}
&=&
\left[ V_{b}+\frac{\alpha _{b}^{2}}{x-\lambda _{b}}\right]^{-1}
+\frac{1}{L^{2}}\sum_{\mathbf{k}}\frac{1}{2\epsilon_{\mathbf{k}}+\hbar \omega_{z}}
\notag \\
&&\hspace{-2cm}
=\frac{\sqrt{2\pi }}{a_{z}^{2}}\left[ \left(U_{p}
+\frac{g_{p}^{2}}{x-\gamma _{p}}\right)^{-1}
-S_{p}(x)+\sigma _{p}(x)\right] .
\label{eq:vbneg}
\end{eqnarray}
Here, $U_{p}=4\pi a_{bg}/a_{z}$, $g_{p}^{2}=\mu _{co}WU_{p}/(\hbar \omega
_{z})$, and $\gamma _{p}=\mu _{co}(B-B_{0})/(\hbar \omega _{z})$ are 3D
dimensionless physical parameters, where $a_{bg}$ is the background scattering
length, $\mu _{co}$ is the difference in magnetic moments between the two
channels, $W$ is the resonance width, and $B_{0}$ is the resonance point.
The functions in Eq. (\ref{eq:vbneg}) take the form
\begin{eqnarray}
S_{p}(x) &=&\frac{-1}{4\sqrt{2}\pi }\int_{0}^{\infty } ds
\left[ \frac{\Gamma(s-x/2)}{\Gamma (s+1/2-x/2)}-\frac{1}{\sqrt{s}}\right],
\label{eq:sp} \\
\sigma_{p}(x) &=&\frac{\ln \left\vert x\right\vert }{4\pi \sqrt{2\pi}},
\label{eq:sigma}
\end{eqnarray}
where $\Gamma (x)$ is the gamma function.
\begin{figure}[htbp]
\includegraphics[width=8.5cm]{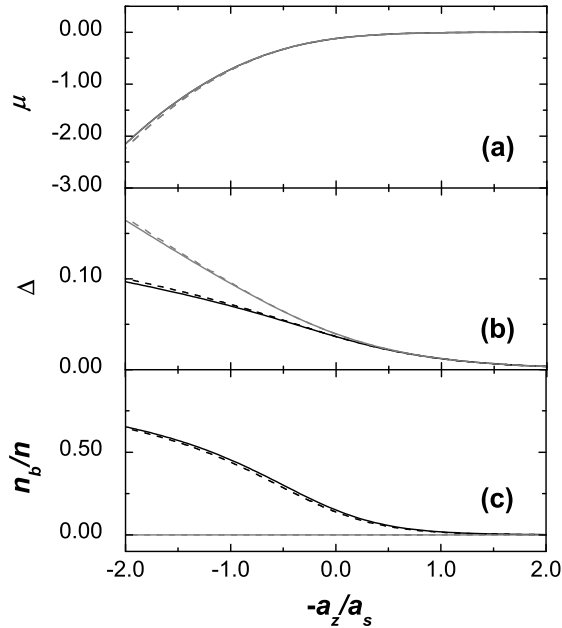}
\caption{The BCS-BEC crossover behavior of a uniform quasi-2D Fermi
gas at zero temperature, showing (a) the chemical potential
$\protect\mu $, (b) the gap $\Delta $, both in unit of $\hbar \omega
_{z}$, and (c) the dressed-molecule fraction $n_{b}/n$. Notice that
the results for $^{6}$Li (solid) and those for $^{40}$K (dashed)
almost coincide as plotted as functions of $a_{z}/a_{s}$, indicating
a universal behavior around the resonance point. Furthermore,
significant difference between model 1 (gray) and model 2 (black)
can be observed in (b) and (c), which shows that model 1 is
oversimplified at unitarity and on the BEC side of the resonance.
The parameters used in these plots are $\omega_{z}=2\pi \times 62$
kHz, and $n a_{z}^{2}=0.001$.} \label{fig:homogeneous}
\end{figure}

Using this model 2 Hamiltonian, we first consider a uniform quasi-2D
Fermi gas with a fixed number density $n$, where the inhomogeneity
in the radial plane is neglected. In this case, the gap and number
equations (\ref{eq:gap2}) and (\ref{eq:number2}) need to be solved
self-consistently for a given magnetic field. A typical set of
results for both $^{6}$Li and $^{40}$K are shown in
Fig.~\ref{fig:homogeneous}, indicates a smooth crossover from the
BCS (right) to the BEC (left) regimes. Here, results obtained from
model 2 (black) are compared with those from model 1 (gray). In this
figure and the following calculation, we use the parameters
$a_{bg}=-1405a_{0}$, $W=300$ G, $\mu_{co}=2\mu_{B}$ for $^{6}$Li,
and $a_{bg}=174a_{0}$, $W=7.8$ G, $\mu_{co}=1.68\mu_{B}$ for
$^{40}$K, where $a_{0}$ and $\mu _{B}$ are Bohr radius and Bohr
magneton, respectively.
\begin{figure}[tbp]
\includegraphics[width=8.5cm]{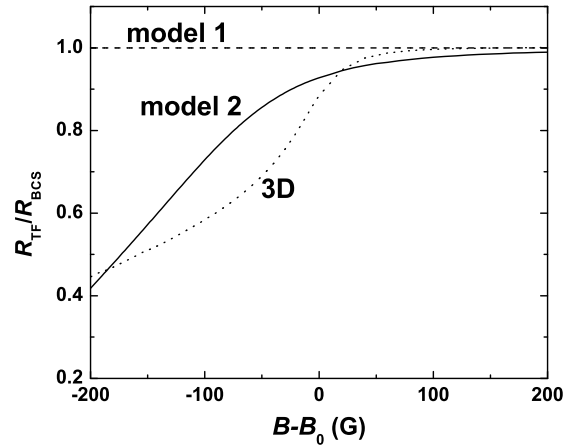}
\caption{The Thomas-Fermi cloud size of a quasi-2D Fermi gas of
$^{6}$Li over a wide BCS-BEC crossover region. Here, results from
model 2 (solid) are compared with those from model 1 (dashed). All
curves are normalized to the cloud size of a noninteracting Fermi
gas $R_{\mathrm{BCS}}$. Notice that the results of model 2 recover
the correct pictures in the BCS and BEC limits, in clear contrast to
the model 1 prediction of a flat line. Parameters used for these two
plots are $\omega _{z}=2 \pi \times 62$ kHz, $\omega _{\perp}=2\pi
\times 20$ Hz, and the total particle number $N=10^{4}$. For
reference, the results for an isotropic 3D Fermi gas with the same
total particle number is also plotted (dotted), where a
single-channel model and a two-channel model are both incorporated
to give indistinguishable predictions. } \label{fig:rtf}
\end{figure}

There are two major points that need to be emphasized in Fig.~\ref{fig:homogeneous}.
First, when plotted as functions of the inverse of 3D scattering length $a_{z}/a_{s}$,
the results for $^{6}$Li (solid) and $^{40}$K (dashed) are very close,
manifesting the near resonance universal behavior. Second, the results from model 1
and model 2 are significantly different, especially on the BEC side of the resonance.
In particular, the dressed-molecule fraction in model 2 is already sizable ($\sim 0.16$)
at unitarity, and becomes dominant on the BEC side of the resonance
(see Fig.~\ref{fig:homogeneous}c). This result provides another signature of
inadequacy of model 1, where the dressed-molecule population is always
assumed to be negligible.

Next, we impose a radial harmonic trap $U(\mathbf{r})$ and calculate
the Thomas-Fermi cloud size for a fixed number of particles in the
trap $N=\int 2\pi n(r)rdr$, as shown in Fig.~\ref{fig:rtf}. The most
important feature of Fig.~\ref{fig:rtf} is that the cloud size given
by model 2 (solid) is no longer a constant as predicted by model 1
(dashed). On the contrary, by crossing the Feshbach resonance, the
cloud size decreases from the limiting value $R_{\mathrm{BCS}}$ of a
noninteracting Fermi gas in the BCS limit, and approaches to the 3D
results (dotted) in the BEC limit. This trend successfully recovers
the corresponding physics in both the BCS and the BEC limits. In
addition, we also find that for a given number of particles in the
trap, the curve trend is insensitive to the radial trapping
frequency $\omega_{\perp}$ within the experimentally accessible
region. (The $\omega _{\perp }=2\pi\times 10$ Hz and $2 \pi \times
50$ Hz results, not shown, coincide with the $2 \pi \times 20$ Hz
line and are hardly distinguishable within the figure resolution.)
Considering the fact that there is a scaling relation between
$\omega _{\perp }$ and $N$ such that the physics is only determined
by $N(\omega _{\perp }/\omega_{z})^{2}$, this insensitivity with
respect to the radial trapping frequency suggests that the
experimental measurement has a rather wide range of tolerance on the
number of atoms.
\begin{figure}[t]
\includegraphics[width=8.5cm]{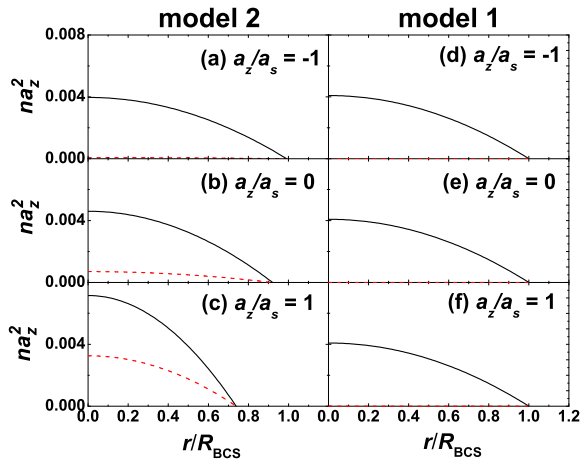}
\caption{(Color online) The in-trap number density (the solid lines)
and dressed-molecule fraction (the dashed lines) distribution along
the radial direction of a quasi-2D Fermi gas of $^{6}$Li, obtained
from model 2 (a-c) and model 1 (d-f). The top panels correspond to
the case of $a_{z}/a_{s}=-1$ (BCS side), the middle panels to the
case of $a_{z}/a_{s}=0$ (unitarity), and the bottom panels to the
case of $a_{z}/a_{s}=1$ (BEC side). The parameters are $\omega
_{z}=2\pi \times 62$ kHz, $\omega _{\perp }=2\pi \times 20$ Hz, and
$N=10^{4}$.} \label{fig:intrap}
\end{figure}

In Fig.~\ref{fig:intrap} we show the number density and the dressed-molecule
fraction distribution along the radial direction for various values of
$a_{z}/a_{s}$. A typical case in the BCS regime is shown in the top panel of
Fig.~\ref{fig:intrap}, where the dressed-molecule fraction is vanishingly
small, and model 1 and model 2 predict similar cloud sizes and number
density distributions. The middle panel shows the case at unitarity.
As compared with model 1, notice that the cloud is squeezed in model 2 and the
dressed-molecule fraction increases to a sizable value. The bottom panel
shows a typical case in the BEC regime, where the cloud is squeezed further
in model 2 as the dressed-molecule fraction becomes significant. Notice that
the results of model 2 successfully describes the BCS-BEC crossover, in
clear contrast to the outcome of model 1.

\section{conclusion}
\label{sec:conclusion}

In summary, we have considered in this paper the BCS-BEC crossover
of a quasi-2D Fermi gas across a wide Feshbach resonance. We
analyze two effective Hamiltonians and compare predictions of zero
temperature cloud size and number density distribution in the
radial plane within a mean-field approach and local density
approximation. Using model 1 with renormalizd atom-atom
interaction, we show that the cloud size remains a constant value
through the entire BCS-BEC crossover region, which is inconsistent
with the picture of a weakly interacting Bose gas of fermion pairs
in the BEC limit. On the other hand, model 2 with renormalized
interaction between atoms and dressed molecules predicts the
correct trend of cloud size variation. Based on this qualitative
comparison, it is likely that the inclusion of dressed
molecules~\cite{kestner-07,duan-05} is essential to describe the
BCS-BEC crossover in quasi low dimensions.

\begin{acknowledgements}

We thank Yinmei Liu for many helpful discussions, and J. P.
Kestner for discussion and thoroughly reading of the manuscript.
This work is supported under the MURI program and under ARO Award
W911NF0710576 with funds from the DARPA OLE Program.

\end{acknowledgements}


\end{document}